# PROOF-OF-REPUTATION: AN ALTERNATIVE CONSENSUS MECHANISM FOR BLOCKCHAIN SYSTEMS

Oladotun Aluko[1] and Anton Kolonin[2]

[1]Novosibirsk State University, Novosibirsk, Russia
[2]Aigents Group, Novosibirsk, Russia

*ABSTRACT*

*Blockchains combine other technologies, such as cryptography, networking, and incentive mechanisms, to enable the creation, validation, and recording of transactions between participating nodes. A consensus algorithm is used in a blockchain system to determine the shared state among distributed nodes. An important component underlying any blockchain-based system is its consensus mechanism, which principally determines the performance and security of the overall system. As the nature of peer-to-peer(P2P) networks is open and dynamic, the security risk within that environment is greatly increased mostly because nodes can join and leave the network at will. Thus, it is important to have a system that can check against malicious behaviour. In this work, we propose a reputation-based consensus mechanism for blockchain-based systems, Proof-of-Reputation(PoR) where the nodes with the highest reputation values eventually become part of a consensus group that determines the state of the blockchain.*

## 1. INTRODUCTION

Distributed consensus among nodes that are geographically distributed has been a widely studied research topic in distributed systems. However, with the introduction of blockchain, it has gotten even more attention because blockchains are a type of distributed system. Most blockchain-based systems aimed at various application domains with a variety of unique requirements have introduced a corresponding consensus mechanism tailored to their specific uses. As a result, several consensus algorithms with varying properties and capabilities have emerged.

A blockchain system, at its core, is a distributed system that uses a consensus algorithm to determine the shared state among distributed nodes. This shared state, known as a chain, is a public or private record of all transactions or digital events that have been created and shared among participating nodes [1]. The primary goal of any blockchain-based system is to maintain a live decentralized transaction ledger while defending against malicious Byzantine actors who may attempt to game the system. The reliability and integrity of the entire blockchain system are heavily reliant on the consensus model used. The applicability of any consensus mechanism is based on three key properties: safety, liveness, and fault tolerance [2].

Aside from appropriate integrity criteria with which transactions are verified, the successful operation of a blockchain system is also dependent on many other key aspects such as the correctness of technical protocols, strong cryptographic mechanisms for managing participant identity and checking transaction integrity, and an incentive to motivate nodes who participate in the community while maintaining an honest behavior over time. Proof-based consensus mechanisms are the most common approach to consensus in blockchain systems. The basic idea behind a proof-based consensus algorithm is that, among the many nodes that join the network, the node that performs sufficient testing is given the right to add a new block to the chain and is rewarded. [3]. A large number of these methods are still vulnerable to the games of the participants on the network. PoW (Proof of Work) consensus algorithm, which powers the Bitcoin cryptocurrency, is the most popular where each participant in the system votes by the total amount of computing power that the participant controls at the time of the vote. The obvious disadvantage of this approach is that anyone with the most computing power can essentially take over a significant portion of the system. In addition, it has been known to consume a significant amount of resources [4].

As peer-to-peer (P2P) networks are open and dynamic in nature, the security risk within that environment is greatly increased, owing to the fact that nodes can join and leave the network at will. As a result, it is critical to have a system in place that can detect malicious behavior. Using community-based reputations is one way to reduce the risks associated with this type of open community. Historically, reputation systems have been employed to facilitate trust between entities [5].

The goal of this work is to create an alternative consensus mechanism for blockchain networks. This consensus mechanism is based on the reputation of network participants. It uses the interaction of nodes in the network over time to determine the amount of reputation associated with each node in the network. It then uses the reputation to determine a set of consensus nodes responsible for maintaining the network's shared state, as well as the calculation of new reputation values as interactions between nodes in the network progress over time.

The main contributions of this work are described as follows:

- First, in our reputation-based consensus mechanism, a node's reputation is calculated by blending together a normalized set of ratings and the corresponding reputation values of the node providing the rating at a given point in time, rather than simply the value of the direct rating given by other nodes. A node's behavior influences its overall reputation value.
- Second, the following principles underpin our reputation-based consensus mechanism: 1) The liquid nature of the reputation values. The reputation value computed for a node is based on the reputation value of the node providing the rating. 2) The temporal scoping of reputation, so that reputation values accumulated by members in the past contribute less to the current reputation value. 3) The openness of all reputation values to all members of the community so that audits can be performed.
- Third, we use a side chain to store all node reputation values, eliminating the need for a third party to manage reputation.
- Finally, we create an experimental implementation and evaluate its performance in terms of system security and throughput.

## 2. REVIEW OF RELATED WORK

### 2.1. Consensus Algorithms

Blockchain relies heavily on consensus and it has been researched for over three decades [6], [7]. Consensus protocols have historically been used to reach an agreement on a shared among a group of distributed nodes. Because designing a consensus protocol is a difficult task, it is common to make assumptions under which the protocol is proven to function properly. These assumptions eventually influence the characteristics of the consensus protocol. In fact, most Blockchain Technology applications typically roll out their corresponding consensus algorithm to fit the specific use case for the Technology [8], [9], [10].

Proof-of-Work (PoW) is by far the most widely used consensus mechanism for blockchains introduced by Bitcoin [2], [11]. In PoW nodes acting as miners vote by the amount of computing power they possess by trying to solve a computational challenge. The first node to solve the challenge validates and adds a new block of transactions to the blockchain, earning a reward. However, the generation of blocks generally requires the use of massive computational power and introduces delay for block confirmation, resulting in low efficiency and transaction throughput.

Proof-of-Stake (PoS) was proposed as an alternative to PoW. Nodes that want to participate in the block creation process with PoS must prove ownership of a certain amount of coins. In addition, they must lock a certain amount of currency, known as stake, in order to participate in the block creation process. Delegated-Proof-of-Stake (DPoS) is a variant of PoS in which miners are elected by other nodes. The stake of the nodes are used as the weighting parameter for votes.

## 2.2. Reputation Systems

The statistical estimate of a node's trustworthiness from the perspective of other nodes is known as reputation. Its predictive power is an indicator of a node's future behavior in a network of interacting nodes. According to the system specification, a good reputation correlates to good behavior and a bad reputation correlates to bad behavior. Reputation is based on ratings received with respect to a particular interaction. Nodes would typically rate other nodes on the basis of this interaction. Specifically, a rating is a judgment from a node (origin) to another node(target) in a scope. Reputation Systems are systems that aggregate these ratings in some way. There are three parts to a Reputation System, the formulation, calculation and dissemination. The formulation describes the mathematical or statistical model and a set of inputs for the assessment of reputation values. The two main parts are: the reputation measure and the mathematical model used to aggregate the ratings. The calculation deals with the design and implementation of the algorithm for the assessment of reputation and finally, the dissemination is about the storage and distribution of reputation in the network. The two popular approaches for managing dissemination is either through a centralised authority or through decentralised nodes in a network of nodes. Furthermore, reputation can be measured using discrete or continuous values. Different methods have been earlier proposed to this effect.

Josang et al. [12], proposed the Beta Reputation System (BRS). In the scheme, ratings are either positive or negative for a trustee which are thereafter considered as two events in a beta probability distribution. The provider's reputation is then calculated as the expected value of the positive rating in the future derived by substituting the number of positive and negative ratings into the beta probability density function. The scheme also uses a forgetting factor to ensure that older ratings have a lower weight.

In the statistical approach proposed by Weng et al. [13], the credibility model is based on statistical approaches. The scheme records the testimonies of all witnesses for each trustee and uses that as a witness profile stored locally by every truster. The credibility of each witness is thereafter evaluated on the basis of its history of success.

Other approaches include those based on fuzzy logic [14], [15], bayesian systems [16], [17], [18] and subjective logic [19], [20].

## 2.3. Reputation-Based Consensus Algorithms

Reputation has been defined as a quantity derived from the underlying social network which is globally visible to all members of the network [21], [22], [23]. Historically, reputation systems have been known as a means of harnessing some form of reputation data. They function by facilitating the collection, aggregation, and distribution of data about a specific entity. This data can thereafter be used to characterize and predict that entity's future actions [24], [25]. Essentially, users within a network can decide who they will trust and to what extent by referring to reputation data. A reputation system, in addition to the foregoing, is a socially corrective mechanism, as the incentive of positive reputation and the disincentive of negative reputation will generally encourage good behaviour in the long run. After a reputation system collects reputation data, it can be shared among users, who can then use it to evaluate other users before making decisions about intended or future interactions, without ever having previously interacted. Examples of the practical application of this system can be found on eCommerce websites like Amazon or eBay where reputation attributed to a seller is influenced by ratings through previous transactions. Another use case is in government where a country like China incentives the behavior of the citizens through a social credit score system. Earlier works proposed possibilities of applying a reputation-based model to distributed computing. One such was described by [26]. The downside was that the approach was not completely decentralized.

Recent studies have introduced reputation systems into the blockchain space to improve efficiency and reliability. [27] proposed a reputation-based consensus mechanism for peer-to-peer networks where reputation serves as the incentive for good behaviour and the node with the highest reputation gets to publish a new block. At the end of each interaction between two nodes, feedback is generated by the service requester and broadcast to the entire network. On reaching the set threshold, nodes start to calculate a ranking list after which the node with the highest-ranking reputation value publishes the new block and other nodes verify the integrity of the newly published block. [28] also proposed a reputation-based consensus mechanism based on the proof-of-work consensus algorithm. In their approach, a miner's voting power is given by its reputation. The reputation for each miner is computed based on the total amount of valid work a miner has contributed and also the regularity of that contribution over a given period. [29] proposed the Blockchain Reputation-Based Consensus(BRBC) mechanism in which a node in the network must have a reputation score higher than a set threshold to be able to publish a new block. Also, a judge is randomly selected that is responsible for updating node reputation values. However, none of these address the behaviour of a node on a transactional basis as it interacts with other nodes in the network.

## 3. BACKGROUND AND SYSTEM CONTEXT

### 3.1. Distributed Systems

In a distributed system, heterogeneous nodes independently run computational tasks and communicate information via message passing. These types of systems typically consist of large numbers of participating nodes. The behaviour of these types of systems become hard to predict as the size of the nodes increase. A fundamental problem associated with distributed systems is that of agreeing on a common state shared across the entire network. The reason is that large distributed systems bring with them some form of inherent non determinism — unpredictable events like out-of-order inputs, message delays, the sudden failure of components, or in extreme cases the unethical actions of faulty or malicious nodes opposed to the goals of the system as a whole.

Most existing agreement protocols are built on the back of some assumptions related to the particular use case for which the protocol will be used. One of the underlying assumptions about these protocols has to do with the network type. Dwork et al [30] categorised these as: synchronous, partially synchronous and asynchronous. In this work, we allude to a couple of network structure principles from [30]. Some of them are:

*Definition 1*: A strongly synchronous network is one in which there exists a known fixed upper bound, $\delta$, such that every message sent at $t$ arrives at another point in the network at most $t + \delta$
*Definition 2*: A partially synchronous network is one in which there exists a fixed upper bound, $\delta$, on a message's delay and one of the following holds:

1. $\delta$ always holds, but is unpredictable.
2. $\delta$ is known, but only holds starting at some unknown time.

Furthermore, the following properties capture the essence of agreement in the network
*Definition 3*: If a correct node outputs a value denoted as *V*, then some other node proposed *V*.
*Definition 4*: If a correct node outputs a value denoted as *V* then all correct nodes output the value *V*.

*Definition 5*: If all correct nodes initiated the protocol then, eventually, all correct nodes output some value *V* within a finite time period *t*.

Definitions 3, 4 and 5 are formally referred to as Non-triviality, Agreement and Liveness properties respectively.

## 3.2. Cryptographic Primitives

### 3.2.1. Hash Functions

We use very secure hash functions generated with the Elliptic Curve Cryptography to generate all types of encryption needed. This method is both efficient and performant. It takes any arbitrarily long string and converts it to a fixed-length string. The main characteristics of this hash function is that for a given message, it is easy to compute the hash; but given the hash, it is difficult to compute the message. Hash functions that demonstrate this property are referred to as one-way hash functions.

### 3.2.2. Digital Signatures

Digital signatures enable the verification of a message. This verification ensures that a message originated from a particular node. Digital signatures are usually a public key, denoted by $pk_i$, used for verifying the signature of node *i* and a corresponding secret key, denoted by $sk_i$, used by node *i* for signing a message. This process is referred to as encryption. To prevent another node or a malicious node from impersonation, a node should not reveal this key. The purpose of the public key is to verify a signature generated by a secret key belonging to a node in the network. A digital signature scheme usually consists of three distinct parts:

1. a key generation algorithm,
2. a signing algorithm, and
3. a verification algorithm.

The key generation algorithm is used to generate a new set of key pairs. The main property required from a digital signature scheme is that of security. Without knowledge of a secret key, it is infeasible to find a string that passes the verification algorithm.

### 3.2.3. Event System

The concept of an event here is similar to traditional distributed algorithms where each event is defined as a unit of computation across a set of nodes. An event can be described as a tuple where $p_i$ is the node at which the event occurs, timestamp representing the exact time the event occurred, and information about the event. This definition is based on the more general formalism from distributed systems where each node of a distributed computation is a state machine and the computation causes a change of state.

### 3.2.4. Information Propagation

Information propagation here deals with two main aspects: the first is about propagating information from a group of nodes to all the other nodes in the network. The other aspect is the propagation of information between two interacting nodes. The goal of information propagation is to provide every node in the network with the most recent information. This information propagation is the essence of communication in a distributed setting. Nodes in the network communicate by transmitting and receiving messages. A Message is therefore a unit of communication. A more general definition is that a message is just a finite sequence of bits. In a distributed environment where information is constantly flowing because of the interactions, there needs to be:

1. an efficient mechanism for distributing this information i.e publish and
2. an efficient mechanism to allow nodes to receive these specific kinds of information i.e subscribe

Specifically, in our scheme, we use a Pub-Sub pattern as the medium for information propagation

### 3.2.5. Leader Election

A leader is a member of a set that is distinguished to perform some special task among other nodes. The leader election problem is the problem of choosing a leader from a set of possible candidates. The fundamental assumption is that every node in the network has a unique identity and also possesses some unique characteristics which might be used as the election criteria. The importance of electing a leader is as a result of the fact that there might be a need to conduct some additional computation for which there is the need to have a coordinator for such computations. It is usually a common practice that the leader is elected on the basis of some competence with regard to the system's operating context. Examples might be the use of compute power as basis for election or the use of memory.

### 3.2.6. Voting

The goal of voting is to aggregate individual preferences among a group into a single group ranking. Weighted voting is a voting method that affords certain nodes more privileges on the outcome of the vote than other nodes based on some specific context for which the voting system is deployed. Weighted voting is a form of social choice system originating from the domain of cooperative game theory [31]. They essentially model decision making processes in which a set of voters make decisions about the outcome of an issue. Each voter is allocated a numeric weight, and the decision is carried out only if the sum of weights of voters in favour of it meets or exceeds some specific given threshold, called the quota.

### 3.3. System Overview and Threat Model

We assume a strongly synchronized system where nodes are connected by a network over which information is propagated through the exchange of messages. We first present the nodes in the network, then we present the adversarial model. In addition, we assume that the network is untrustworthy and unreliable, which means messages in the network can be delayed, duplicated or lost in some cases. Furthermore, the nodes are heterogeneous and failure at a node does not cause the failure of another node.

#### 3.3.1. Nodes

We assume that there's a social community that affords nodes the ability to vote about different aspects of the system. Each node $i$ is identified by a public key $pk_i$ similar to a wallet in regular blockchains like Bitcoin. This public key has a corresponding secret key $sk_i$ with which it can use to append its signature to transactions. Each transaction group is made up of two nodes $[i,j]$. Node $j$ gives the rating while node $i$ is the recipient of this rating usually with respect to an interaction between them. The transaction is said to be completed only after it is appended to the blockchain.

#### 3.3.2. Adversary

Nodes can be in one of two broad categories: honest nodes and faulty nodes. Honest nodes are nodes that follow the protocol precisely, and faulty nodes which do not follow the protocol. Furthermore, a faulty node can either be the result of unresponsiveness due to a crash or lack of information in which case they are referred to as sleepy nodes or Byzantine faulty nodes which are nodes that act arbitrarily. We denote the total number of nodes in the network as $N$, the number of faulty nodes as $f$ which is a sum of both Byzantine nodes and sleepy nodes. We restrict the adversary to control at most faulty nodes, where $3f + 1 \leq N$. We also assume the presence of an adversary that can control all the faulty nodes and can coordinate their behavior.

## 4. METHODOLOGY

### 4.1. Consensus Mechanism

#### 4.1.1. Consensus Group

In our scheme, we assume $N$ nodes in the network, an individual node is represented as $p_i, i \in N$. Network nodes perform computation in rounds, during which a node sends messages, receives them, and then performs some local computation on the received message [32]. Each node $i$ is identified by a key pair, $pk_i$ which is the public key and $sk_i$ is the corresponding secret key. At the end of every interaction between nodes, rating values are generated with respect to the service. A node will typically function in one of two ways: as the recipient node or as the rater node for the specific interaction. When a rater node gives a rating, it broadcasts the transaction details to the entire network. We formally denote this interaction where rater node is $i$ and node $j$ is the recipient node as a tuple:

$$Transaction = (E_{sk_i}, pk_j, r) \quad (1)$$

$$\{r : 0 < r < 1\}$$

where *pk<sub>j</sub>* is the public key of the recipient node, *r* is the rating given by the rater node which is a value between 0 and 1, and *Esk<sub>i</sub>* is the encrypted transaction data signed using the rater node's secret key. Transactions generated between nodes for a round *k* are added to a list of pending transactions waiting to be appended to the chain during the consensus phase.

At the start of every consensus round, consensus group members need to be selected and added into a consensus group. We denote this consensus group for a round *k* as $G_k$. Members of the consensus group are chosen from the nodes with the highest reputation values, with collective reputation scores that exceed 50% of the total reputation values of the network. With this approach, the size of $G_k$ will vary depending on the reputation distribution in the network. A node that is part of the consensus group is denoted as:

$$pk_i \in G_k \quad (2)$$

To proceed, a new leader $L_k$ for the round *k* is selected. After the leader for the round *k* is selected, it serves the following functions:

- Packaging all valid transactions from the list of pending transactions to a *Block<sub>k</sub>*
- Calculating the new reputation values for all network nodes for *Reputation<sub>k</sub>* the round *k* using data from transactions in the transaction list
- Broadcasting the commit message to the consensus group $G_k$

---
**Algorithm 1:** Node Transaction
**Result:** information propagated in the network as a message
input: *rating* and *recipient*;
1: transaction = createTransaction(*rating, recipient*);
2: transactionPool.setTransaction(*transaction*);
3: broadcastTransaction(*transaction*);
---

Algorithm 1: Node Transaction

### 4.1.2. Leader Selection

We use a random function to select the leader for the consensus group for each round $G_k$. By doing this, there is no deterministic guarantee for which node will be selected as leader for the consensus round *k*. As such, all nodes that have been selected as members of the consensus group for the round *k* have equal chance of being selected. The leader's public key is broadcasted to the consensus group before the start of the consensus phase. The leader $L_k$ packages all the transactions $T_i$ for a recent time window in a specific order and adds them into a block. Afterwards, the leader sends a commit message to the consensus group $G_k$. This message contains the newly packaged *Block<sub>k</sub>*, the leader's public key $pk_L$, reputation list for the round *k*, and also a hash of the *Block<sub>k</sub>* generated using the leader's secret key:

$$< Block_k, pk_L, Hash(Block_k), ReputationList > \quad (3)$$

```
Algorithm 2: Leader Selection
Result: Leader L for the round k
input: set of highest ranking nodes G;
1: leader = random(G);
2: broadcastLeader(leader);
```

Algorithm 2: Leader Selection

### 4.1.3. Block Publication

After a commit message is sent by the leader $L_k$ to the consensus group $G_k$, the consensus group determines the final block to be published for the current round $k$. Each node $pk_i \in G_k$ checks the commit message sent by the leader $L_k$ which contains both the $Block_k$ and $Hash(Block_k)$. First, the node checks the pkL in the commit message to see if it matches the pkL that was broadcasted upon the leader selection earlier; otherwise, it can ignore the commit message. It then proceeds to check the integrity of the hash using the pkL. Afterwards, it checks the validity of the transactions within the $Block_k$. If this process completes, it sends a new commit message back to the consensus group $G_k$. This process continues with every node in the consensus group $G_k$. Upon completion, each node $pk_i$ that successfully verifies the $Block_k$ and the *ReputationList* sends a verification commit back to the consensus group $G_k$.

$$< VERIFY, Block_k, Hash(Block_k), ReputationList > \quad (4)$$

The consensus group waits until at least a certain amount of consensus members sends in this message. This message constitutes a consensus group vote. We formalize this using a social choice function. For a set of nodes in the consensus group for round $G_k$, each node has an associated weight $w$ assigned which is equivalent to its reputation value from the previous round $k - 1$. During the consensus process, there's a minimum quota that has to be reached for decisions to be made. We set this quota at two-thirds of the total weight in the consensus group:

$$d(G_k) = \begin{cases} 1 & \sum_{i \in G_k} w_i > q \\ 0 & \text{otherwise} \end{cases}$$

where $d(G_k)$ represents the decision of the consensus group. Whenever the $d(G_k)$ is 1, it means consensus for round $k$ has been reached.

### 4.2. Reputation System

The reputation of a node is determined by an analysis of the ratings it has received from others in the past. Based on previous interactions, these ratings reflect the level of trust that other nodes have in a specific node. In general, reputation-based systems rely on feedback to evaluate a node. This feedback is generally in terms of the amount of satisfaction a node receives by interacting with another node in the network. According to [33], when considering reputation information, the source of information as well as the context must be considered. We define the reputation principles for the approach adapted from [34] as follows:

- The liquid nature of the reputation values. The reputation value computed for a node is based on the reputation value of the node providing the rating.

- The temporal scoping of reputation so that reputation values collected by members in the past are less contributing to the current reputation value.
- The openness of all reputation values to all members in the community so that audits can be performed.

Let $s_i$ denote the reputation for a recipient node *i*. All nodes in the network start with a default reputation value determined on system initialization. During node interactions, a node's reputation value is determined by the liquid rank algorithm [35]. This approach can be used as a predictive metric to evaluate a node's behaviour. For each round, a node can receive multiple unique ratings

$$s_{i1...n} = \{s_{i1}, ..., s_{in}\} \quad (5)$$

where the range of $s_i$ is *[0, 1]*. Values $s_i$ are then normalised as follows

$$S_{i,n} = \frac{S_{i,n} - min_i(S_{i,n})}{max_i(S_{i,n}) - min_i(S_{i,n})} \quad (6)$$

We slightly modify the normalization of the rating values to prevent null values from the set of ratings as follows

$$S_{i,n} = \frac{(S_{i,n} - min_i(S_{i,n})) + 1}{(max_i(S_{i,n}) - min_i(S_{i,n})) + 1} \quad (7)$$

Furthermore, we define the ratings matrix *S* to be *[s_{ij}]*. After each round, these ratings will be generated for all nodes in the network. To compute new reputation values for a node for the round *k*, we blend these ratings with the rater reputation values from the previous round *k − 1*. We denote this as

$$P = \vec{S} * \vec{R} \quad (8)$$

where $\vec{S}$ = *[s_{ij}]* and $\vec{R}$ =*[r_{in}]*. $r_{in}$ corresponds to the rater node providing the rating.

To compute the reputation value for the round *k*, we then blend the initial node's reputation value with the current rank generated from the ratings.

$$R_{i,k+1} = \alpha * P + (\alpha - 1) * R_k \quad (9)$$

where *α* is a constant determined on system initialization. The value is set between 0 and 1. It determines what portion of the equation to give more priority to. If the value is set closer to 1, it means that the newly generated reputation value will give more priority to the ratings *P* and less priority to the previously generated reputation value. This is what we want as this aligns with the reputation principles stated earlier. It helps to reduce the impact of nodes that change behaviour over time. Further, to prevent reputation values from hopping, we clamp the values using a sigmoid function as follows

$$R'_{i,k+1} = \frac{R_{i,k+1}}{\sqrt{1+(R_{i,k+1})^2}} \qquad (10)$$

Most existing reputation mechanisms rely on a central server to store, manage, and, in some cases, distribute reputation values among network nodes. We do not need a central server in our approach, and the reputation value for each node in the network is managed via a reputation side chain linked to the main transaction chain. The structure of the reputation chain is such that it has a header which contains meta information about a specific block and then the reputation values for all the network nodes:

$$ReputationBlock_i = (ReputationBlockHeader_i, ReputationList_i) \qquad (11)$$

The *ReputationList$_i$* contains a list of all network nodes with their associated reputation values for the most recent round. This serves as a lookup data structure for future uses. The *ReputationBlock$_i$* as well as the transaction block use the standard blockchain block structure with a hash of all the transactions for the round, a previous hash, timestamp and transactions.

A new reputation block is created along with a normal transaction block during the consensus phase. So for a consensus round k, a *Block$_k$* which is added to the transaction chain corresponds to a *ReputationBlock$_k$* which is added to the reputation chain. As stated in section 4, part of the duties of the consensus group is to validate the reputation calculation generated by the leader $L_k$. After the consensus is reached, the leader broadcasts the new *ReputationBlock$_k$* to the entire network and as such the reputation value of all the nodes in the network is visible to all other nodes.

## 5. EXPERIMENTS AND RESULTS

For our prototypes, we built an experimental protocol that implements the protocol. Thereafter, the nodes were deployed on AWS EC2 remote server running on 16GB RAM with Amazon's t3 processor. In the experiment, we set up 1,000 nodes. We used a default initial reputation value of 0.2. We set the value of α to 0.6, the effect of that is that we give priority to recently generated reputation values.

We impose a round trip latency of 200ms to simulate the effect of a Wide Area Network. While this is unlikely to occur in practice, the average of network delays across the entire network will average out to a close enough value.

In terms of the throughput, Figure 1 shows how the throughput values change as we vary the number of network nodes from 500 to 1,000 nodes. As the network size increases, so does the throughput because as more nodes join the network, more messages are being transferred in the network.

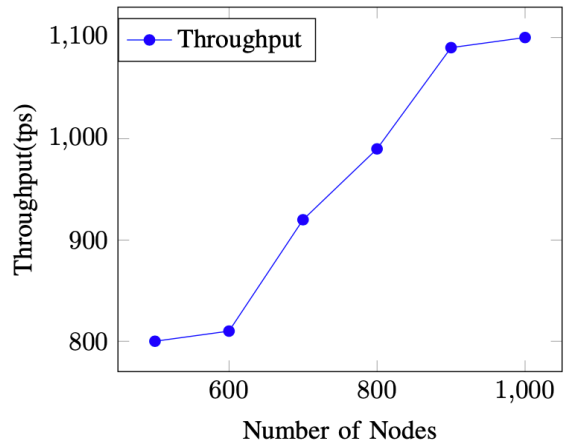

Figure 1: Throughput vs Number of network nodes

Also, we vary the number of transactions in a single block, we vary them between 100 and 500 to measure the average time it takes for a new block to be produced. Figure 2 shows that it takes more time for a new block to be produced as the transactions in a block increase. This is so because more transactions are now in a single block and so it takes more time for those transactions to be processed.

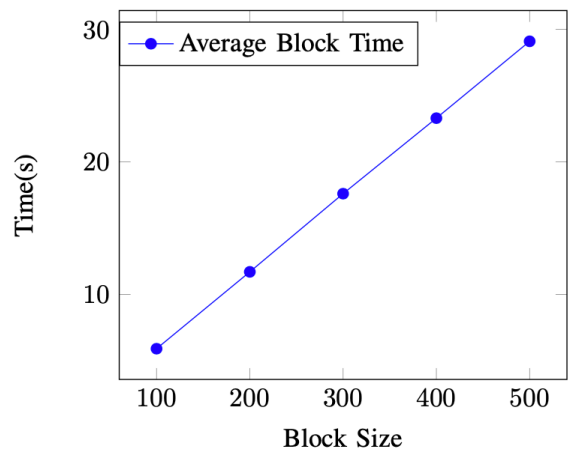

Figure 2: Average Block Time as the number of transactions in a single Block is varied

In our final experiment, we measure the consensus time against the block size for each block. We observed that when the block size is relatively small around 100 transactions in each block, consensus takes about 2 seconds. As we increase the block size, so does the consensus time increase as well.

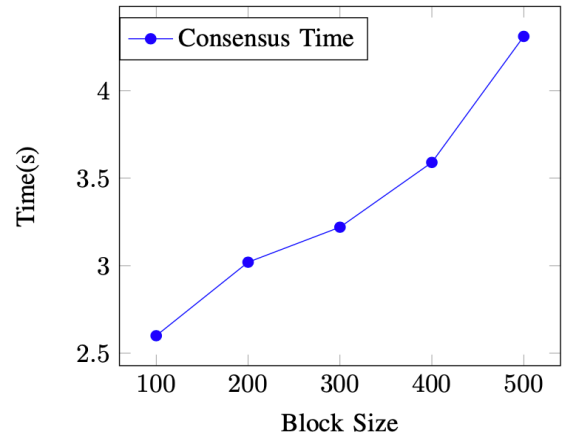

Figure 3: Consensus Time as the number of transactions in a single Block is varied

Table 1 shows the performance of our scheme against other existing consensus mechanisms.

Under certain conditions when the number of participating nodes is increased, our scheme can achieve up to 1,100 transactions per second.

Table 1. Comparison with other consensus mechanisms

| Consensus Mechanism | Throughput(TPS) |
|---|---|
| Proof-of-Work | 7 |
| Proof-of-Stake | 60 |
| Proof-of-Reputation(Baseline) | 800 |
| Proof-of-Burn | 854 |
| Proof-of-Reputation | 1,100 |

## 6. ANALYSIS AND SPECIFIC ATTACK VECTORS

Our consensus mechanism satisfies three important properties: safety, liveness, and fairness. The consensus mechanism is generally safe as long as the bound on the byzantine nodes holds. The consensus mechanism is also live. Based on the network type, a new block will eventually be added to the chain even if there is some delay on the propagation of information. Finally, the consensus mechanism is fair. Here, we refer to fairness in three aspects: fairness in the selection of the consensus group members, fairness in the election of the consensus round leader and fairness in terms of openness of all transactions and reputation values to all participants of the network.

### 6.1. Safety

In the worst case scenario, we consider the existence of a byzantine node with high enough reputation value which has successfully been included in the consensus group $G$ for a round. The consensus remains safe if:

1. The attacker does not control more than f nodes in the consensus group
2. The collective reputation values of the nodes compromised by the attacker is less than 1/3 of $G$

The system can continue as long as (1) or (2) do not hold. Furthermore, due to transmission delays that may cause the system to lose some of its liveness, a node in the consensus group does not have to wait for every other node in the consensus group but only a subset of the nodes where the collective reputation values is greater than 2/3 of the total consensus for $G$.

## 6.2. Liveness

In a general sense, liveness guarantees that all nodes that participate in the consensus can finally reach a consensus result. In the network assumption, there is an upper bound which represents the limit at which a message will eventually be delivered even after it might have been delayed.

*Definition 6*: A protocol is live if and only if for every honest node $i$ and a finite time $t_0$:

1. there is a time period $t_1 > t_0$ where the node $i$ will have received a newly committed block
2. there is a time period $t_2 > t_0$ where every other node will have received the newly committed block

## 6.3. Fairness

The consensus mechanism achieves fairness in three aspects: fairness in the selection of the consensus group members, fairness in the election of the consensus round leader and fairness in terms of openness of all transactions and reputation values to all participants of the network. Unlike in PoW where the consensus is centered around the compute power, the compute power has no effect on the security in our scheme. In addition, simply having a high reputation value is not enough because of the nature of the consensus protocol.

In most cases, a malicious node in the network exhibiting some unpredictable behaviour that deviates from the goal of the system will receive low ratings from other nodes in the network. In this case, the node will never even have a high enough reputation to make it into the consensus group $G$ therefore, it cannot successfully launch a meaningful attack that would disrupt the security of the system.

## 6.4. Attack Resilience

On initialisation, the system bootstraps with all nodes having a default reputation value. At this stage, the system is at its most vulnerable. The reason for this is that the reputation value is equal across the entire network is equal and there is no way to distinguish between honest and byzantine nodes especially for cases where there's a byzantine node in the network. In addition, reputation values need time to grow. During these early stages also, after a few rounds, the variation in reputation values for the highest ranking nodes and the lowest ranking nodes are not significant enough to explicitly distinguish the orientation of the nodes in the network.

Assume a consensus group with 15 nodes having a combined reputation of over 50% of the entire network. The attacker can function in two ways in the consensus group. It may be the case that it is:

1. It is selected as the consensus group leader $L$
2. It is simply a consensus group member

In the first case, as the consensus leader, the attacker node can attempt to compromise the global shared state or global reputation by increasing its reputation which would consequently mean it would be part of future reputation rounds or try to reduce the reputation value of other nodes or both at the same time. On the basis of our random leader selection, in the presence of a malicious node, the probability *Pr(i)* of that node becoming the leader is 1/15. As the size of the network grows and consequently the size of the consensus group increases, the value of *Pr(i)* reduces more and more making it more difficult for a malicious node to attempt to compromise the system security.

Furthermore, during the voting process, the dynamic voting weight value is assigned to a node on the basis of its reputation value. The implication of this is that the higher the reputation value, the higher the vote weight.

For a malicious node, even if it possesses the highest reputation value in the consensus group, it still is not able to compromise the system security as the consensus can only be reached with a coalition of nodes with reputation value greater than 2/3 of *G*.

Under certain conditions as shown in the graph below, where a significant amount of consensus nodes are controlled by a malicious node, the security can significantly come under threat. Where the number of consensus nodes is 15, *f = 4*. Specifically where the combined reputation value of the *f* nodes exceeds the quota, these coalition of nodes can successfully launch a successful attack. However, as the consensus group size grows, the minimum required *f* nodes increases and so a malicious node effectively has to control more nodes for which we argue that for larger networks, this will be difficult to accomplish.

### 6.5. Specific Attack Vectors

#### 6.5.1. Selfish Mining Attacks

Selfish Mining attacks [36], [37], [38] is a mining strategy where a group of miners collude to exert power over the entire blockchain in order to increase their revenue. In selfish mining attacks, two groups exist side-by-side: an honest group of miners following the standard protocol and a colluding group that follows the selfish mining strategy. The selfish miners mine blocks while keeping them secret, they continue this process until the fork created from the main chain is longer than the main chain. In our approach, since blocks are not mined based solely on the compute power a node possesses or a group of nodes collectively possess, this kind of attack is impossible. Furthermore, there is no way for a node to know the nodes that will be involved in the consensus for a round or which node will be selected as the leader for that round.

#### 6.5.2. Eclipse Attacks

An eclipse attack [39] happens whenever a node in the network is occluded from the rest of the network. Most of the external contact for that node is controlled by the malicious node that launched the attack. This attack is a serious threat to any blockchain. In the case of our approach, the effect of this type of attack is only noticeable if an attacker is able to simultaneously isolate multiple consensus group members which is highly unlikely.

#### 6.5.3. Flash Attacks

Flash attacks [40] happen whenever an attacker can purchase or rent compute power for a short period with the intention of using this compute power to its advantage. This type of attack is only feasible with network types like PoW that require the use of compute power. In our approach, an attacker with a sufficiently large amount of compute power cannot simply launch an attack on the basis of its compute power.

Table 2. Attack Resilience

| Attacks | PoW(Bitcoin) | PBFT(ByzCoin) | PoR |
|---|---|---|---|
| Liveness | ✓ | ✗ | ✓ |
| Flash Attacks | ✗ | ✗ | ✓ |
| Selfish Mining Attacks | ✗ | ✗ | ✓ |
| Eclipse Attacks | ✗ | ✓ | ✓ |

# 7. CONCLUSIONS

In this work, we proposed a reputation-based consensus mechanism for distributed ledger systems. The consensus scheme uses a social choice function where the weight of nodes that are responsible for consensus is equivalent to the reputation value for that node. In addition, our approach uses the liquid rank algorithm where the reputation of a node is calculated by blending the normalized ratings by other nodes in the network for a given period with the reputation values of the nodes giving the ratings. Finally, we built an experimental prototype to show the potential of this approach. We remark that there are several other parts of this system which can be improved and are left to future work.

International Journal of Network Security & Its Applications (IJNSA) Vol.13, No.4, July 2021

## AUTHORS

**Oladotun Aluko** received his BSc(2017) in Computer Science and Engineering from Obafemi Awolowo University, Nigeria. He is currently working on his MSc in Big Data Analytics and Artificial Intelligence at Novosibirsk State University, Novosibirsk, Russia. His research interests are in Distributed Computing, Blockchain Technology, Machine Learning and Cloud Databases.

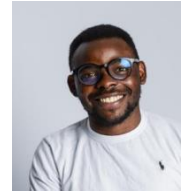

**Anton Kolonin** received his PhD in 1998 after he independently developed a software-algorithmic complex for processing geophysical data, introduced into production in many CIS countries. He has also participated as a leader or lead architect in many projects to develop algorithms and software, including those related to the use of AI, including the recognition of static text, moving objects, music, extracting information from texts and identifying events on financial markets – in Russian and foreign companies. Since 2017, he has also been a software architect for AI and blockchain in the Singularity NET project, leading projects on unsupervised language learning and reputation systems.

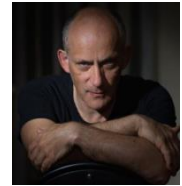